\documentclass[twocolumn,footinbib,floatfix,showpacs,preprintnumbers,amsmath,amssymb,pra,superscriptaddress,longbibliography,notitlepage]{revtex4-1}

\usepackage{epsfig,amscd,wasysym}
\usepackage{times}
\usepackage{amsmath}
\usepackage{amsfonts}
\usepackage{amssymb}
\usepackage{graphicx}
\usepackage{color}
\usepackage{accents}
\usepackage[margin=20mm]{geometry}
\usepackage{array,mathtools,xcolor,physics,xr-hyper,hyperref,natbib}
\usepackage{dsfont}
\usepackage{float,graphicx,color,curves,wrapfig,physics}
\usepackage{amsfonts,physics}
\usepackage{amsmath,gensymb,enumerate}
\usepackage{enumitem}
\usepackage{lipsum}
\usepackage[caption=false]{subfig}

\hypersetup{citecolor=violet}
\hypersetup{linkcolor=red}
\hypersetup{urlcolor=blue}
\hypersetup{colorlinks=true}

\newcommand{\be}{\begin{equation}}
\newcommand{\ee}{\end{equation}}
\newcommand{\bea}{\begin{eqnarray}}
\newcommand{\eea}{\end{eqnarray}}
\newcommand{\non}{\nonumber}

\newcommand{\phd}{\phantom\dagger}

\newcommand{\nin}{\noindent}

\makeatletter % the lines below make the double bra and double ket symbols
\newsavebox{\@brx}
\newcommand{\llangle}[1][]{\savebox{\@brx}{\(\m@th{#1\langle}\)}%
  \mathopen{\copy\@brx\kern-0.5\wd\@brx\usebox{\@brx}}}
\newcommand{\rrangle}[1][]{\savebox{\@brx}{\(\m@th{#1\rangle}\)}%
  \mathclose{\copy\@brx\kern-0.5\wd\@brx\usebox{\@brx}}}
\makeatother

\newlength{\dhatheight} % lines below make the double-hat symbol

\newcommand{\qed}{\nobreak \ifvmode \relax \else
      \ifdim\lastskip<1.5em \hskip-\lastskip
      \hskip1.5em plus0em minus0.5em \fi \nobreak
      \vrule height0.75em width0.5em depth0.25em\fi}

\usepackage{enumitem}
\usepackage{lipsum}
\usepackage{dsfont}

\usepackage{array}
\usepackage{colortbl}  % For \arrayrulecolor

\usepackage{booktabs}
\newcommand{\One}{1\kern-4.5pt1}

\newcommand{\bal}{\begin{align}}
\newcommand{\eal}{\end{align}}
\newcommand{\bas}{\begin{align*}}
\newcommand{\eas}{\end{align*}}

\begin{document}

\title{Measurement Induced Dynamics and Trace Preserving Replica Cutoffs }
	\author{Graham Kells }% \includegraphics[scale=0.066]{./ORCID}\href{https://orcid.org/0000-0003-3008-8691}}
	\email{graham.kells@mu.ie}
	\affiliation{Maynooth University, Maynooth, Co. Kildare, Ireland.}
	\affiliation{Dublin Institute for Advanced Studies, 10 Burlington Road, Dublin 4, Ireland.}	

\date{\today}

\begin{abstract}
We present a general methodology for addressing the infinite hierarchy problem that arises in measurement-induced dynamics of replicated quantum systems. Our approach introduces trace-preserving replica cutoffs using tomographic-like techniques to estimate higher-order replica states from lower ones, ensuring that partial trace reduction properties are rigorously maintained. This guarantees that the dynamics of single-replica systems correctly reduce to standard Lindblad evolution. By systematically mapping information between replica spaces of different orders, we characterise null spaces under partial trace operations and outline efficient algorithmic approaches to enforce positivity.  Importantly, it is demonstrated that pre-calculated stochastic Gaussian ensembles of free fermion states provide an effective and computationally efficient means to stabilise the replica hierarchy, even in the presence of interactions.  Numerical tests on small interacting fermionic systems illustrate the effectiveness and practicality of our approach, showing precise agreement with trajectory methods while providing significantly better statistical convergence.
\end{abstract}

%\pacs{74.78.Na, 74.20.Rp, 03.67.Lx, 73.63.Nm, 05.30.Ch}
\maketitle

Recent discoveries have shown that continuous measurement can drive quantum many-body systems through entanglement phase transitions \cite{Li2018, Skinner2019, Chan2019, Li2019, Szyniszewski2019, Cao2019, Bao2020, Turkeshi2020, Choi2020, Zabalo2020, Jian2020, Chen2020, Tang2020, Nahum2021, Sang2021, Turkeshi2021, Lavasani2021, Buchhold2021, Alberton2021, Block2022, Agrawal2022, Piccitto2022, Kells2023, Fava2023, Poboiko2023, Leung2023, Fava2024, Fisher2023review, Potter2022review}. These transitions separate distinct dynamical phases: an area-law phase where quantum information remains localized, and a volume-law phase characterized by macroscopic information scrambling. This phenomenon can also be viewed from the perspective of purification dynamics  \cite{Gullans2020, Ippoliti2021} but the core idea is the same: the competition between unitary evolution (which generates entanglement) and measurement (which extracts information and can reduce entanglement) leads to qualitatively different behaviours in the long-time dynamics of quantum states.  One key challenge in studying these transitions is that they cannot be detected through simple measurement averages but require access to the full measurement distribution statistics or entanglement measures across ensembles of pure states. 

A standard approach to accessing these quantities is through quantum trajectories  \cite{Dalibard1992, Carmichael1993, Plenio1998, Gardiner2004, Wiseman2009, Daley2014} , which provides an efficient computational framework for simulating individual measurement realisations. However, trajectory methods have some limitations. To use them numerically for large system sizes, we need to restrict dynamics to classes of problems that are classically simulable - specifically Clifford circuits and matchgate/free-fermion dynamics.  The trajectory approach also presents a challenge for experimental implementations. The need to accumulate statistics over many trajectories leads to an exponential scaling in resources - a challenge known as the postselection problem.  The fast-growing literature on this includes brute force  \cite{Noel2022,Koh2023} , dual unitaries \cite{Ippoliti2021b,GoogleQAI2023},  branching circuit architectures \cite{Feng2025}, and other hybrid classical-quantum methods \cite{McGinley2024,Li2023,Garratt2024}

\begin{figure}[h]
\centering
\includegraphics[width=0.47\textwidth]{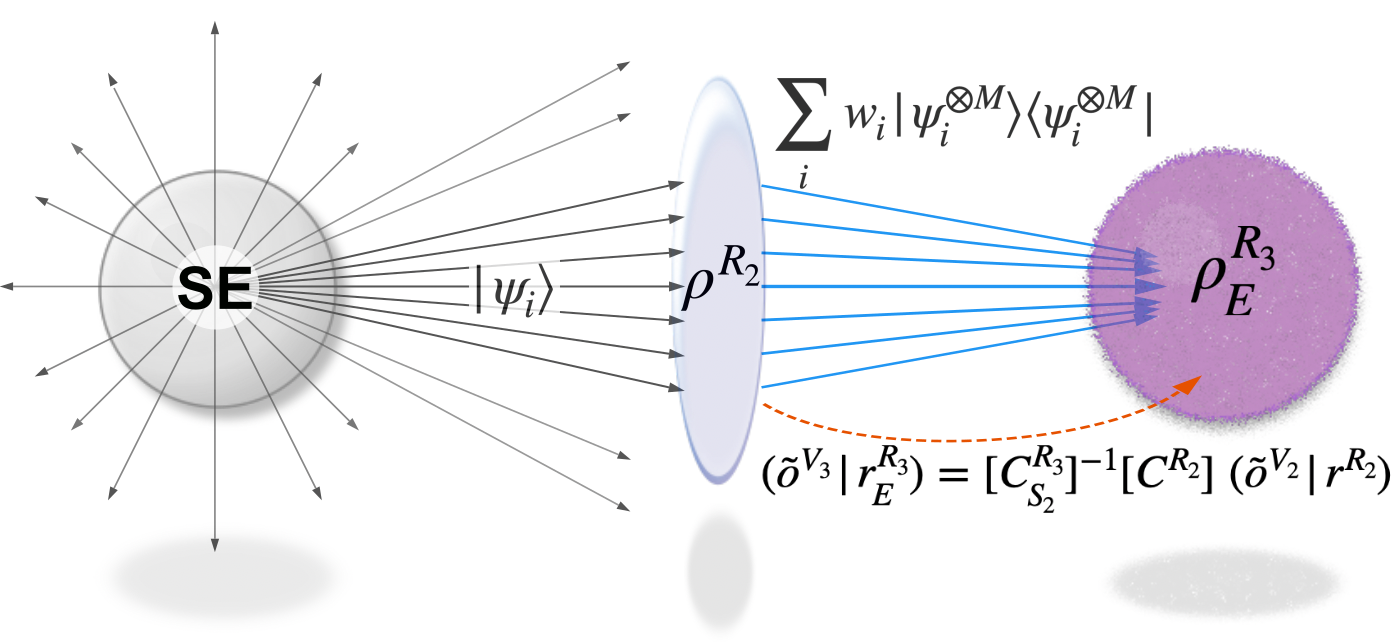}
\caption{Higher replica estimation: Data from a lower replica state (in this schematic $\rho^{R_2}$) is used to construct a positive semi-definite higher replica estimate from a Stochastic Ensemble (SE). Precise real space data from the lower replica $\rho^{R_2}$ is then exactly transposed onto this higher order estimate - ensuring correct behaviour of this replica state ($\rho_E^{R_3}$) under partial trace. }
\label{fig:ReplicaEstimate}
\end{figure}

Here we explore an alternative approach that uses replicated quantum systems, where measurement outcomes become correlated between copies, leading to non-linear master equations \cite{Buchhold2021}. This deterministic method provides direct access to partial purities - (which is closely related to averaged Renyi entanglement entropies) from which we can detect and study entanglement transitions. However, the replica approach introduces its own challenges: an infinite hierarchy of coupled equations between different replica orders. While this hierarchy can be handled theoretically for infinite replicas \cite{Fava2023, Poboiko2023, Fava2024, Poboiko2025},  general numerical implementations require a truncation scheme. In this paper, we present a method for performing replica cutoffs that preserves the crucial property of partial trace reduction. Our approach uses tomographic-like techniques to estimate higher-order replica states from lower ones, ensuring that higher-order replica states always reduce correctly to lower-order ones under partial traces (see Figure \ref{fig:ReplicaEstimate}). When implemented within a master equation framework, our method guarantees that the dynamics reduce properly to the original single-copy Lindblad equation. % (see Figure \ref{fig:RT_method}).
%\begin{figure}[h]
%\centering
%\includegraphics[width=0.5\textwidth]{RT_method}
%\caption{Replica cut-off schematic: (a) For a single copy, trajectory average and Lindblad give the same average results (b) Master equations on the 2-replica level require information from 3 and 4 replica states, which in turn need information from higher order replicas. (c) The 2-replica can be estimated  from data obtained on the single replica. (d) A 2-replica master equation cut-off can be made by estimating 3- and 4- replicas from the 2-replica data while ensuring the higher replicas are physically reasonable. }
%\label{fig:RT_method}
%\end{figure}
It also provides two further key advances: First, it offers a systematic way to map information between replica spaces of different orders. Second, it enables direct construction and categorisation of null spaces under the partial trace operation. These null spaces are crucial because higher-order replica estimates are not guaranteed to be positive semi-definite (PSD) and thus may not represent physical states. We demonstrate how to enforce the PSD constraint through optimisation in these null spaces, providing a practical numerical implementation through hybridisation with stochastic ensembles.

\vspace{1mm}
\nin {\em Methods:} The measurement-induced dynamics of quantum systems can be described by the stochastic Schrödinger equation (SSE):
\be
\dd \ket{\psi_t} = -i \dd{t} \left[ \hat{H} - \frac{i\gamma}{2} \sum_i \hat{M}^2_{i,t} \right] \ket{\psi_t} +\sum_i \dd W_i \hat{M}_{i,t} \ket{\psi_t} \non
\ee
where $\hat{M}_{i,t} = \hat{O}_i - \langle \hat{O}_i\rangle_t$ are measurement operators constructed from local operators $\hat{O}_i$, and $\dd W_i$ represents Gaussian white noise with $\overline{\dd W_i} = 0$ and $\overline{\dd W_i \dd W_j} = \gamma \dd t \delta_{ij}$.

The methodology that we introduce transcends specific Hamiltonians, but we will focus our examples on tight-binding fermionic Hamiltonians with the possibility of breaking  Gaussianity via density-density interactions. The $L$-site lattice Hamiltonian  for this is given by:
%\begin{equation}
%\label{eq:Hamiltonian}
%\begin{split}
%	H_0 &=  - \mu \sum_{x=1}^{L}  (c^\dagger_{x} c_{x }-1/2) \\
%	-&  w \sum_{x=1}^{L} c^\dagger_{x}c_{x+1} + h.c + \Delta \sum_{x=1}^{L} c^\dagger_{x} c^\dagger_{x+1} + h.c. ,
%\end{split}
%\end{equation}
\begin{equation}
\label{eq:Hamiltonian}
\non H =  -\sum_{x=1}^{L} c^\dagger_{x}c_{x+1} + \text{h.c} +V \sum_{x=1}^{L-1}  (n_x - \frac{1}{2}) (n_{x+1}-\frac{1}{2})
\end{equation}
where $c^{(\dagger)}_x$ represent fermion (creation) annihilation operators, $n_x= c^\dagger_x c^{\phd}_x $, $w$ is the kinetic hopping amplitude, and $V$ the density-density interaction strength. The $O_i$ will be related to the lattice fermion number $O_i = 1-2 n_i$. Some recent work on measurement-induced dynamics in such interacting models can be found in  Refs. \cite{Poboiko2025,Cecile2024,Lumina2024}.

\begin{figure*}
\centering
\includegraphics[width=1\textwidth]{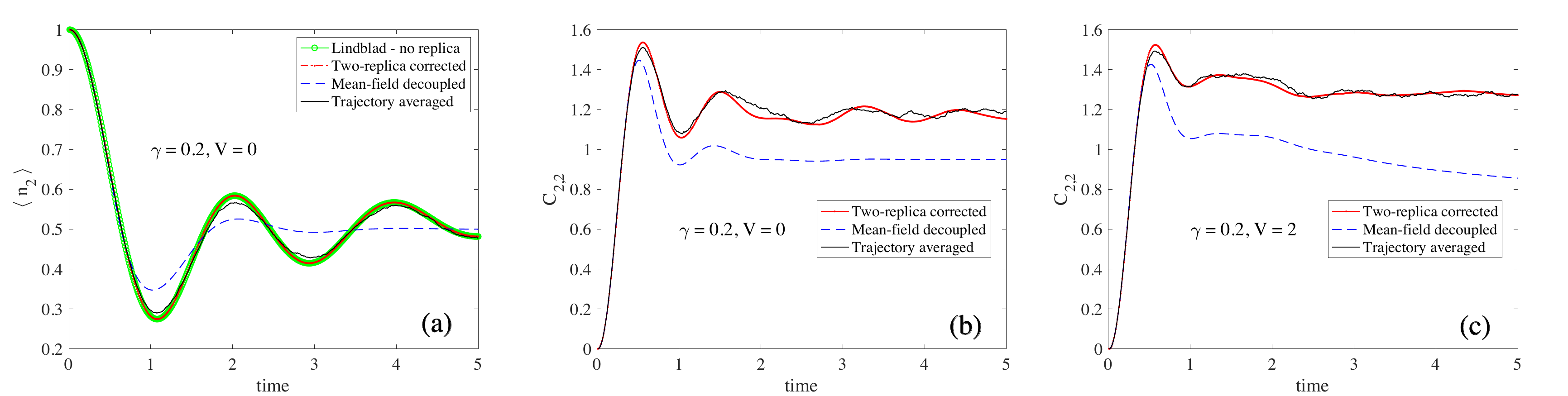}
\caption{(a) The partial trace-preserving replica cut-off ensures dynamics always reduces to Lindbladian evolution when traced back to a single replica (b) Trajectory hybridised approach leads to a stable $R_2$ corrected master equation with excellent convergence properties. (c) A single pre-calculated stochastic ensemble can also be used to stabilise the 2-replica master equation. In this example, a Gaussian Stochastic Ensemble is used to cut off and stabilise interacting ($V=0.4$) two-replica dynamical calculations.  All of the figures are for measurement strength  $\gamma=0.5$ and system size $L=4$.}
\label{fig:Examplestopfigure}
\end{figure*}

\nin{\em Replica density matrices - construction and properties}
For any single trajectory labeled by $c$, we construct the conditional density matrix $\rho^{(c)}_t = \dyad*{\psi_t^{(c)}}{\psi_t^{(c)}}$. The physical density matrix is obtained by averaging over trajectories:
$
\rho_t \equiv \rho^{R_1}_t = \frac{1}{N_c} \sum_{c=1}^{N_c} \rho^{c}_t \equiv \overline{\rho^{c}_t}
$
The replica approach extends this by considering tensor products of conditional matrices before averaging:
\be
\rho^{R_n}_t = \frac{1}{N_c} \sum_c (\rho^{c}_t)^{\otimes n} \equiv \overline{(\rho^{c}_t)^{\otimes n}}
\ee
This construction provides direct access to higher moments of measurement statistics through cross-correlation between replicas. For example, while $\rho^{R_1}_t$ gives only average values $\langle \hat{O}\rangle$, the two-replica density matrix $\rho^{R_2}_t$ encodes correlations like $\overline{\langle \hat{O}\rangle^2}$ that are essential for detecting entanglement transitions. This construction also possesses two fundamental properties that are crucial for our methodology: 

\nin 1. {\it Partial trace reduction}: Tracing out any replica reduces the state to a lower-order replica: $\Trace_{(i)} \rho^{R_n}_t = \rho^{R_{n-1}}_t  $
%\be
%\non
%\Trace_{(i)} \rho^{R_n}_t = \rho^{R_{n-1}}_t  
%\ee

\nin 2. {\it Permutation symmetry}: Expectation values are invariant under replica exchange. For operators $\hat{A}$, $\hat{B}$ acting on single replicas:
\be
\begin{split}
\Tr \hat{A}^{(i)} \rho^{R_n} = \Tr \hat{A}^{(j)} \rho^{R_n}  &\\
\Tr \hat{A}^{(i)} \hat{B}^{(j)} \rho^{R_n} = \Tr \hat{A}^{(k)} \hat{B}^{(l)} \rho^{R_n} &
\end{split}
\ee
for any replica indices $i,j,k,l$. These properties enable a systematic method for mapping information between replica spaces of different orders while preserving the essential physical structure.

\vspace{1mm}
\nin{\em Replica master equations and the infinite hierarchy:}
%Starting from the SSE differential equation for the conditional pure-state density matrix as 
%\bea
%\dd \rho^c_t  &=& -i \dd t [\hat{H}, \rho^c_t]   + \sum_i \dd W_i \left\{ \hat{M}_i, \rho^c_t \right\} -   \\ \non  && \frac{\gamma \dd t}{2} \sum_i  \left\{ \hat{M}^2_i , \rho^c_t \right\} + \sum_{ij} \dd W_i \dd W_j \hat{M}_i \rho^c_t \hat{M}_j
%\eea
Buchhold et al. \cite{Buchhold2021} considered the replicated system and computed SSE for the 2-replica density matrix:
\bea
\non
\dd \rho^{R_2}_t &=& \rho^{R_2}_{t+ \dd t} - \rho^{R_2}_t  % = \overline{\rho^c_{t+ \dd t} \otimes  \rho^c_{t+ \dd t}} -\overline{\rho^c_{t} \otimes  \rho^c_{t}} 
= \overline{ \dd \rho^c_{t} \otimes  \rho^c_{t}} + \overline{\rho^c_{t } \otimes  \dd \rho^c_{t}}+ \overline{ \dd \rho^c_{t} \otimes \dd \rho^c_{t}}
\eea
where the first two terms reduce to
\bea
\label{eq:dpall}
\overline{ \dd \rho^c_{t} \otimes  \rho^c_{t}} &=& \dd t \mathcal{L}(\rho_t) \otimes \rho_t \equiv \dd t \mathcal{L}^{(1)} (\rho^{R_2}_t) \\ \non
\overline{\rho^c_{t} \otimes \dd \rho^c_{t} } &=& \dd t  \rho_t  \otimes \mathcal{L}(\rho_t) \equiv \dd t \mathcal{L}^{(2)} (\rho^{R_2}_t).
\eea 
and the last to
\bea
\label{eq:dpdp}
\overline{ \dd \rho^c_{t} \otimes  \dd \rho^c_{t}} 
 &=&   \gamma \dd t\sum_i \left\{ \hat{O}^{(1)}_i ,\left\{ \hat{O}^{(2)}_i , \overline{\rho^c_t \otimes \rho^c_t} \right\}\right\}   \\ && \quad - 2 \gamma \dd t \sum_i  \left \{ \hat{O}^{(1)}_i+\hat{O}^{(2)}_i,\overline{ \langle \hat{O}_i \rangle_t  \rho^c_t \otimes \rho^c_t} \right\} \non \\ && \non \quad + 4 \gamma \dd t \sum_i  \overline{\langle \hat{O}_i \rangle_t^2  \rho^c_t \otimes \rho^c_t} 
\eea
where $\hat{O}_i^{(1)}= \hat{O}_i \otimes I$,  $\hat{O}_i^{(2)}= I\otimes \hat{O}_i  $ are operators acting on different replica subspaces and $ \langle \hat{O}^{ }_i \rangle_t  \equiv \langle \hat{O}^{(1)} _i \rangle_t =\langle \hat{O}^{(2)} _i \rangle_t$ because of the inherent symmetry between replicas. 

The latter two terms cause a problem \cite{Buchhold2021} because it is not possible to fully disentangle statistical correlations between the expectation values $\langle \hat{O} _i \rangle$ and $\rho^c_t \otimes \rho^c_t$.  However,  these terms can be calculated using higher-order replicas :
\be
\begin{split}
\overline{ \langle \hat{O}_i \rangle_t  \rho^c_t \otimes \rho^c_t}  = \Tr_{(3)} \left[ \hat{O}_i^{(3)} \rho_t^{R_3} \right]   \\
\overline{ \langle \hat{O}_i \rangle_t^2  \rho^c_t \otimes \rho^c_t}  = \Tr_{(3,4)} \left[ \hat{O}_i^{(3)} \hat{O}_i^{(4)} \rho_t^{R_4} \right]
\end{split}
\label{eq:T3T4}
\ee
whereby $\rho_t^{R_3}$ and $\rho_t^{R_4}$ are the three and four replica density matrices.  This leads to an apparent problem in that to calculate the 2-replica state $\rho^{R_2}$, you need to also calculate the $\rho^{R_3}$ and $\rho^{R_4}$ states - and to calculate them, you need higher-order replica states and so on - setting off an infinite hierarchy of replica dependency. 

A practical workaround is to use the notion of a mean-field like decoupling such that we replace these terms with
\bea
\overline{ \langle \hat{O}_i \rangle_t \rho^c_t \otimes \rho^c_t } &\rightarrow&  \overline{ \langle \hat{O}_i \rangle_t } \times  \overline{  \rho^{c\phantom |}_t \otimes \rho^{c\phantom |}_t }  = \overline{ \langle \hat{O}_i \rangle_t } \times  \rho^{R_2}  \non  \\  \overline{ \langle \hat{O}_i \rangle_t^2 \rho^c_t \otimes \rho^c_t } &\rightarrow & %\overline{ \langle \hat{O}^{(1)}_i \hat{O}^{(2)}_i  \rangle_t } \times  \overline{  \rho^{c\phantom |}_t \otimes \rho^{c\phantom |}_t } 
    \overline{ \langle \hat{O}^{(1)}_i \hat{O}^{(2)}_i  \rangle_t } \times  \rho^{R_2}    \non
\eea
This allows the master equation hierarchy to be {\em cut off}  or {\em closed} at order two.
%\bea
%\label{eq:dpR2}
%\dd \rho^{R_2}_t  &=& \non  (\mathcal{L}^{(1)} + \mathcal{L}^{(2)} - 4 \gamma \overline{C_t} ) \dd t  \rho_t^{R_2}     \\ &+ & \dd t \gamma \sum_i \left\{ \hat{O}^{(2)}_i - \overline{\langle \hat{O}^{(2)}_i\rangle} ,\left\{  \hat{O}^{(1)}_i - \overline{\langle \hat{O}^{(1)}_i\rangle}, \rho^{R_2}_t]\right\} \right\} \non
%\eea
%where $\overline{C_t} = \sum_i \overline{ \langle \hat{O}^{(1)}_i \hat{O}^{(2)}_i \rangle_t} - \overline{\langle \hat{O}^{(1)}_i \rangle_t } \;\;  \overline{\langle \hat{O}^{(2)}_i \rangle_t }$ has been introduced to preserve the unit trace.  
Although this approximation may be well motivated for many scenarios,  a key problem with this (see \cite{Supplementary}) % Appendix \ref{sect:TP}) 
is that it does not preserve partial traces.  This means that in general, when reduced to a single replica, the dynamics reproduce the expected Lindblad behaviour.

 \vspace{1mm}
\nin{\em Information transfer between replicas of different orders:}
This problem of partial trace preservation does not arise if one uses the original expressions \eqref{eq:T3T4}. However, to preserve lower-order replica dynamics, requiring the higher replica states to be exact is overly restrictive. Indeed, there is significant freedom to modify the higher $\rho^{R_M}$ provided these modifications are null under the partial trace reduction to the next lowest replica state $\rho^{R_{M-1}}$.

A direct approach to systematically transfer information between replica spaces is to represent the replica density matrix in terms of its Hilbert-Schmidt projections onto a complete set of observables. Working with a vectorized notion of the Hilbert-Schmidt inner product 
$\left(A | B\right) \equiv \Tr A^\dagger B /  (\Tr A^\dagger A \times  \Tr B^\dagger B)^{1/2} $
we can express an $M$-replica density matrix  as
%\bea
$| \rho^{R_M} ) = \sum_{j=1}^{N_\text{max}}  (\hat{O}_j  | \rho^{R_M} ) |  \hat{O}_j ) $
%\eea
where $j$ runs over all $N_{\text{max}}=4^{LM}$ orthogonal operators $\hat{O}_j$. When we can only access a subset of replica spaces $N<M$, we cannot determine all weights needed to represent the exact state. Here we can hope to estimate the higher replica density matrix by including the weights we can reliably determine. However, replica density matrices should also be physical - unit trace and positive semidefinite - and this gives us a means to fill in some of the missing information.
%\bea
%|\rho_{E}^{R_M}) &=&  \frac{1}{4^{L M}} \sum_{j=1}^{S_N}  (\hat{O}_j  | \rho^{R_M} ) |  \hat{O}_j )
%\eea
%That is $N_{\text{max}=S_N$ is the number of operators that act nontrivially on a maximum of $N<M$ replicas. 

To formalise this approach, one can first construct a permutation symmetric basis for the replica spaces. These basis vectors denoted $\ket*{V^{R_N}_i}$, reside in the full Hilbert space of the $N$-replica system. The dimension of this symmetric subspace is $D_N={{d+N-1}\choose{N}} $ where $N$ is the number of replicas and $d$ depends on the number of qubits/sites in the bare system and any additional symmetries such as number/parity conservation.  As the density matrices $\rho^{R_N}$  are permutation symmetric, they can be fully projected to the corresponding symmetric subspace  $r^{R_N}_{ij} = \bra*{V^{R_N}_i}  \rho^{R_N}   \ket*{V^{R_N}_j}$. 
%$\rho^{R_N} = P^{R_N} \rho^{R_N}  P^{R_N}$, where $P^{R_N} = \sum_i |V^{R_N}_i \rangle \langle {V^{R_N}_i}|$ and we can define the projected matrix representation

For a set of operators $\hat{O}_i$, we define their projections onto the symmetric subspace as $ \hat{o}^{R_N}_{ij} = \bra*{V^{R_N}_i}  \hat{O}  \ket*{V^{R_N}_j}$. Although the original operators $\hat{O}_i$ form an orthonormal set under the Hilbert-Schmidt inner product, $( \hat{O}_i | \hat{O}_j) = \delta_{i,j}$, their projections onto the symmetric subspace are generally not.  Within a subspace of dimension $D_N$, there are at most $S_N=D_N^2$  independent Hermitian operators. This number can be reduced in the case of high symmetry. For example, in the hopping model described above, at half-filling with no interactions, the dimensionality of independent operators is instead $S_N = (D_N+1)(D_N)/2$.

We now denote with $\hat{O}^{V_N}$ a subset of the $\hat{O}$ operators that, when projected to the symmetric subspace, form a linearly independent set. The projections of these operators are denoted as $\hat{o}^{V_N}$ and we can therefore construct an orthonormalized set $|\tilde{o}_i^{V_N})$ such that 
$(  \tilde{o}^{V_N}_i | \tilde{o}^{V_N}_j) = \delta_{ij}$. If we then define the overlap matrix $C^N_{ij}= (o^{V_N}_i | \tilde{o}^{V_N}_j)$ that relates the non-orthogonal and orthonormal spaces, we can relate real space expectation values to their orthonormalised counterparts:
%\be
$(\tilde{o}^{V_N}_i | r^{R_N}) = \sum_j [C^N]^{-1}_{i,j} (\hat{O}^{V_N}_j |\rho^{R_N})$.
%\ee
As the set $\hat{O}^{V_N}$ that is linearly independent when projected to a symmetric space $\ket*{V^{R_N}}$ will also be linearly independent when projected to a higher replica symmetric space $\ket*{V^{R_{M}}}$ with $M>N$, one can therefore construct mappings between replica spaces of different orders:
\bea
(\tilde{o}_i^{V_M}| r^{R_M}_E) = \sum_{jk}[C_{S_N}^M]^{-1}_{i,j} [C^N]_{j,k} (\tilde{o}_k^{V_N}| r^{R_N})
\eea
where $C_{S_N}^M$ is the matrix containing the first $S_N$ rows and columns of $C^M_{ij}= (o^{V_M}_i | \tilde{o}^{V_M}_j)$. The reconstructed higher-order replica estimate
$
\rho^{R_{M}}_{E} = \sum_{i,j} [r_{E}^{R_{M}}]_{i,j} \ket*{V^{R_{M}}_i} \bra*{V^{R_{M}}_j}.
$
will retain the crucial partial trace property  $\Tr_{(a,b)} \rho^{R_{N+2}}_{E} = \Tr_{(a)} \rho^{R_{N+1}}_{E} = \rho^{R_{N}}.$
%\be
%\label{eq:pt_property}
%\Tr_{(a,b)} \rho^{R_{N+2}}_{E} = \Tr_{(a)} \rho^{R_{N+1}}_{E} = \rho^{R_{N}}.
%\ee
Further mathematical details of this construction are provided in \cite{Supplementary}  where we also outline alternative means to navigate the partial trace null spaces using eigenspaces of lower-rank replicas.
%Appendix \ref{sect:projected_overlap_method}. 
%In Appendix \ref{sect:eigenbasis_method},

\vspace{1mm}
\nin {\em Positivity and partial trace null spaces:}
The estimated states $r_E^{R_M}$ (and equivalently $\rho_E^{R_M}$) accurately encode correlations up to $M^\text{th}$ order but miss higher-moment information. This manifests in the spectrum of these density matrix estimates, which are not generally positive semi-definite (PSD) despite correctly reproducing all lower-order replica properties. The non-positivity is nonetheless useful because it signposts a path back to the PSD manifold where the true higher-order $\rho^{R_M}$ resides. By enforcing the PSD constraint while maintaining the lower moment information obtained from $\rho^{R_N}$, we can make surprisingly accurate estimates of the true higher replica state. Within the PSD manifold, one could optimise further by maximising suitable entropy measures \cite{Jaynes1957a,Jaynes1957b}.

The projected operator basis described above provides a convenient way to explore this replica operator space. It is graded such that projected operators $\tilde{o}_i^{V_M}$ for $i > S_N$ span the null space of the partial trace operation. This means that weights $(\tilde{o}_i^{V_M}|r_E^{R_M})$ for $i > S_N$ can be freely adjusted without affecting the property specified above. % equation \eqref{eq:pt_property}. 
To enforce the PSD constraint, we seek values for these null space coefficients that make the overall density matrix positive semi-definite. Using brute force minimisation, we have had success with semi-definite programming approaches—specifically SDPT3 \cite{SDPT3_a, SDPT3_b} and MOSEK \cite{MOSEK} within the CVX \cite{CVX_a, CVX_b} environment. %However, we eventually encounter scaling issues, particularly when inferring $\rho^{R4}$ for even modest system sizes. Indeed, even for a system of size $L=4$ at half-filling, the dimensions grow rapidly: $S_2 = 231$, $S_3 = 1596$, and $S_4 = 8001$. 

Solving large SDPs at each time step is however prohibitive. A more promising approach, see Figure \ref{fig:ReplicaEstimate}, focuses weighted ensembles of pure states that best fit the lower replica data. For any replica order $M$, we can approximate $\rho^{R_M}$ with an ensemble of weighted pure states:
\begin{equation}
\rho^{R_M} = \sum_i w_i \ket{\psi_i^{\otimes M}}\bra{\psi_i^{\otimes M}},
\end{equation}
with weights $w_i$ chosen to maximize fidelity with the known lower-order ($N<M$) replica state. This ensemble construction automatically guarantees positivity while typically yielding high-entropy approximations, naturally biasing toward physically reasonable higher-order states.

One final step transposes known exact lower replica data onto higher replica estimates using the methodology discussed above. This step forces precise agreement with single replica Lindblad evolution. Although this can weakly break PSD again on higher replica estimates, it nevertheless brings us ever closer to the correct higher replica states. 

For good convergence, proposed pure states must respect the underlying symmetries of the model.  Of course, an excellent ensemble would be one comprised of numerically calculated trajectories themselves. However, the ensembles don't have to be calculated alongside master equation evolution. Alternative strategies include generating ensemble states using the density matrix structure at a given time as a seed state, systematically modifying weights and states from previous timesteps, or pre-calculating a fixed stochastic ensemble of pure states and optimising only weights $w_i$ at each step.  

\vspace{1mm}
\nin {\em Numerical Tests:}
Key numerical metrics are the expectation values $ \langle \hat{n_i} \rangle_t =   \Tr  \hat{n}^{(j)}_i \rho_t^{R_2}$  and the inter replica cross-correlation 
$
C_{ij}(t) =  \Tr \left[ ( \hat{O}^{(1)}_i - \hat{O}^{(2)}_i) \times(\hat{O}^{(1)}_j - \hat{O}^{(2)}_j)  \rho_t^{R_2} \right]
$
This inter-replica correlation - representing moments higher than the mean -  can't be obtained from single-replica GKSL dynamics \cite{Buchhold2021}. %Could also add 2-Renyi entanglement entropy.

Figure \ref{fig:Examplestopfigure} shows several outputs of simulations for measurement rate $\gamma=0.2$. In  (a) the exact agreement of the corrected 2-replica master equation and the Lindblad calculation is shown, along with the inherent discrepancy introduced via the mean-field-like decoupled cut-off.  In  (b) and (c) we show the non-linear correlation $C_{1,2}$ for both non-interacting and interacting scenarios. In this calculation, a fixed random Gaussian ensemble approximates the $\rho^{R_2}$ correlators, forcing the PSD condition on higher-order replicas at each time step. The exact $\rho^{R_2}$ data is then transposed onto the stochastically estimated $\rho^{R_3}$ and $\rho^{R_4}$.  More numerical details are presented \cite{Supplementary}.

\vspace{1mm}
\nin{\em Discussion:} 
A method for constructing partial trace-preserving replica cutoffs is presented that solves a key challenge in the replica framework: truncating the infinite hierarchy of coupled equations while maintaining lower replica consistency.  The framework estimates higher-order replica states from lower ones using an approach that rigorously preserves partial trace reduction properties. This ensures that truncated replica master equations correctly reduce to Lindbladian evolution for single-copy systems. By characterising the null spaces under partial trace operations, we provide a precise mathematical description of the freedom available when constructing higher-order replica estimates, enabling effective approaches for enforcing positivity constraints while preserving lower-order moments.

A notable practical contribution is the demonstration that pre-calculated ensembles of free fermion states can effectively cut off and stabilise the replica hierarchy, even for interacting systems. This eliminates the need to generate system-specific trajectories alongside the master equation evolution, significantly reducing the overhead. Numerical tests on small fermionic systems confirm the method's effectiveness, showing good agreement with traditional trajectory methods while providing better statistical convergence.  An important future direction would be to quantify the behaviour of the corrected master-equation dynamics with respect to the characteristics of the stochastic ensemble. For example, assuming the ensemble is some $t$-design, how does $t$ affect both the convergence properties and stability of the method?

Despite these advances, scaling these methods to larger systems remains challenging. The primary limitation is the exponential scaling of the replicated many-body Hilbert space. One potential avenue would be implementing these trace-preserving cutoffs within tensor network frameworks, particularly MPS representations. These approaches naturally compress the many-body Hilbert space and could provide efficient pathways for locally enforcing positivity constraints through their inherent variational structure.

{\em Acknowledgements} I  thank Alessandro Romito, Dganit Meidan, Tara Kalsi, Chun Leung, Joost Slingerland, Shane Dooley, Luuk Coopmans, Joshuah Heath and Anthony Kiely for helpful discussions on this paper and related topics.

\bibliography{refs2.bib}

\appendix

\pagebreak

\section{Additional numerical examples}

Here we assemble some of the other numerical simulation data that indicates some of the key properties of the method.  

\begin{figure*}[h]
\centering
\includegraphics[width=1\textwidth]{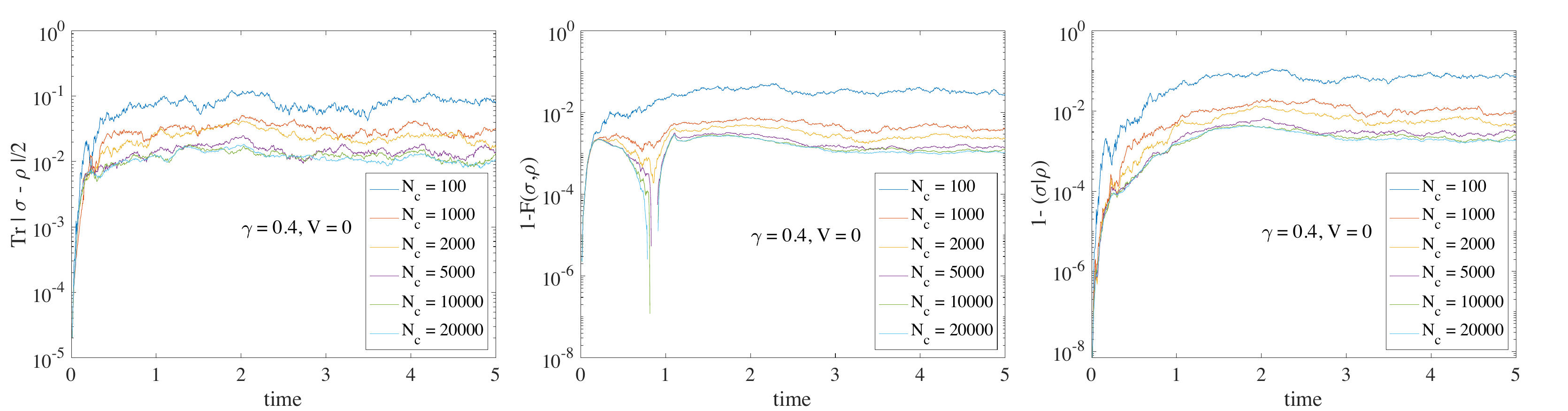}
\caption{Various measures of the difference between the trajectory calculated replica density matrix $\sigma = \overline{\rho^c \otimes \rho^c}$, where $N_c$ is the number of trajectories, and the corrected Master equation density matrix $\rho=\rho^{R_2}$. Here, as in the main text, we use a random Gaussian Ensemble to cut off and stabilise the $\rho^{R_2}$ dynamical calculations.  All of the figures are for measurement strength  $\gamma=0.4$ and system size $L=4$.  Differences tend to saturate around the order of the Trotter step ($\delta t = 0.01$).}
\label{fig:fidelity}
\end{figure*}

In Figure \ref{fig:fidelity} we examine a number of overlap measures to assess how the accuracy of the method. For a fixed Gaussian ensemble of $N_{GE} = 4000$, we compare against the trajectory averaged dynamics for various sample sizes. Convergence of the method is robust to the expected error associated with the Trotterized timestep.   

\begin{figure*}[h]
\centering
\includegraphics[width=0.95\textwidth]{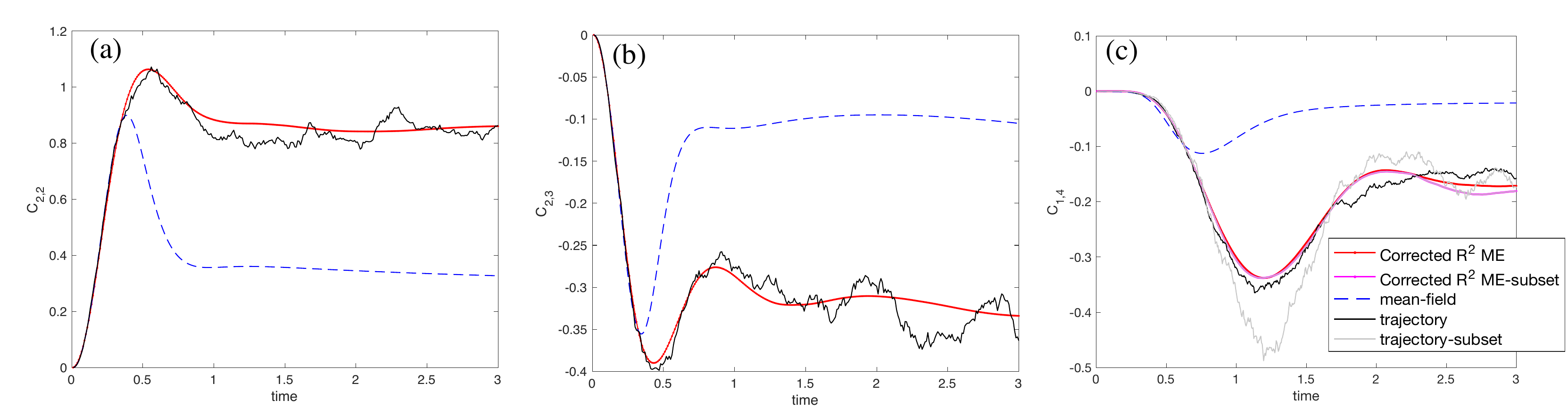}
\caption{(a) and (b) $R^2$ corrected master equation on a system for  $L=4$ and $\gamma=0.5$, and density-density interactions $V=0.4$. Crucially, the same random Gaussian ensemble is used to stabilise the $R^2$ master equation cut-off at each time step. As before, the non-linear correlations match the behaviour of full trajectory approaches. In this graph, we used ensemble sizes of $N=1000$  for the interacting trajectory calculations. (c) Here, we use trajectories calculated in parallel to the master equation to stabilise the cut-off. Again, the non-linear correlations match the behaviour of full trajectory approaches - albeit with significantly better convergence -  here we used $L=4$, $V=0$, $\gamma=0.3$ and ensemble sizes of $N=1000$ and $N=100$ for the respective trajectory subset calculations. }
\label{fig:Ccompare}
\end{figure*}

In Figure \ref{fig:Ccompare} we plot the value of the correlators $C_{ij} (t)$.   In figures (a) and (b) we showcase data obtained using the fixed ensemble method.  In (c) we show how trajectories, calculated in tandem to master equation, can be used to stabilise and estimate non-linear correlations. This hybridized approach can be used to smooth away much of the statistical noise associated with trajectories. 

\begin{figure*}[h]
\centering
\includegraphics[width=0.95\textwidth]{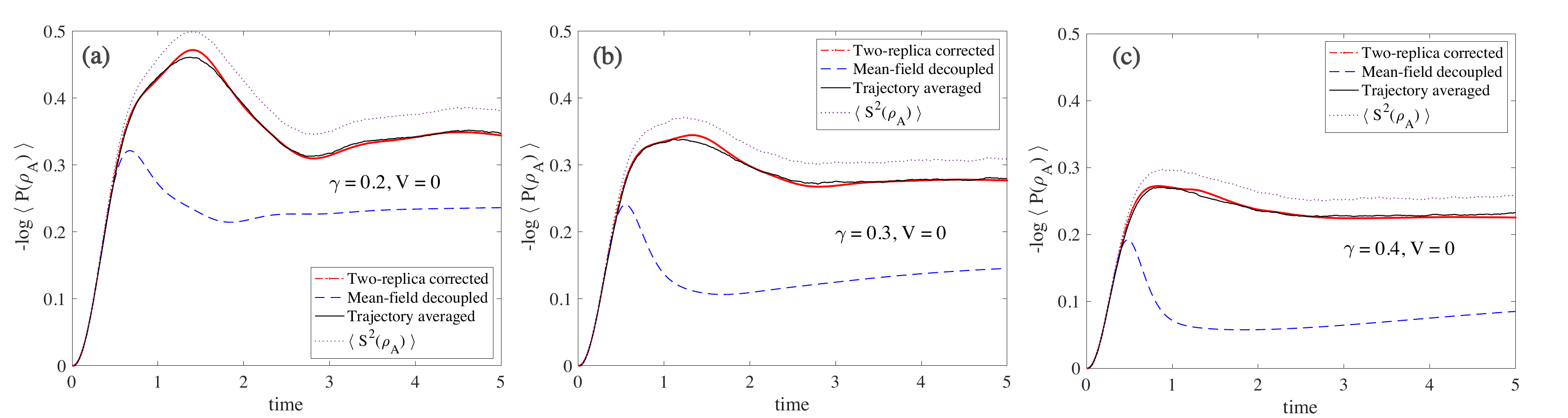}
\caption{The averaged 2nd Renyi Entropy is calculated via trajectory ensembles and compared against $-\log \langle P(\rho_A) \rangle$ calculated via the same ensemble, the corrected master equation method, and the mean-field decoupled master equation.   The 2-replica corrected master equation method is stable and matches well with the trajectory average. }
\label{fig:R2_comparison}
\end{figure*}

One of the key motivations for using the replica methodology is that it provides direct access to  entanglement measures that are averaged over the pure states that make up the ensemble. In Figure \ref{fig:R2_comparison} we plot calculations of the entanglement purity 
$$
\langle P (\rho_A) \rangle =  \frac{1}{N_c} \sum_c  \Tr ( (\rho^{(c)}_A)^2  ) =  \Tr \left( \chi_A \overline{ \rho \otimes \rho} \right),
$$
where  $\rho^{(c)}_A = \Tr_B \rho^{(c)}$ and $\chi_A$ is a swap between replica copies of the Hilbert space $A$,   comparing averages of equal weight ensemble of $N_c$ trajectories with the value obtained from the $\rho^{R_2}$ replica master equation calculation.

Note that the $n^{\text{th}}$ R\'{e}nyi Entropy can written as $$ S_n(\rho_A) = (1/(1-n)) \log( \Tr(\rho_A^n)).$$ 
Therefore, while we can write the averaged $2^{\text{nd}}$ R\'{e}nyi entropy as
$$
\langle S_2(\rho_A) \rangle =  - \frac{1}{N_c} \sum_c \log \left( \Tr(  (\rho^{(c)}_A)^2 ) \right),
$$
with replica states $\rho^{R_2}$ we only have access to 
$$
-\log \langle P (\rho_A) \rangle = - \log \left(\frac{1}{N_c} \sum_c \Tr ( (\rho^{(c)}_A)^2 ) \right),
$$
and so the precise equality between entanglement purity and R\'{e}nyi entanglement entropy only holds for pure states.  The value $-\log \langle P (\rho_A) \rangle$ serves as a strict lower bound for the $2^{\text{nd}}$ R\'{e}nyi entropy in the case of mixed states.

\begin{figure*}
\centering
\includegraphics[width=0.95\textwidth]{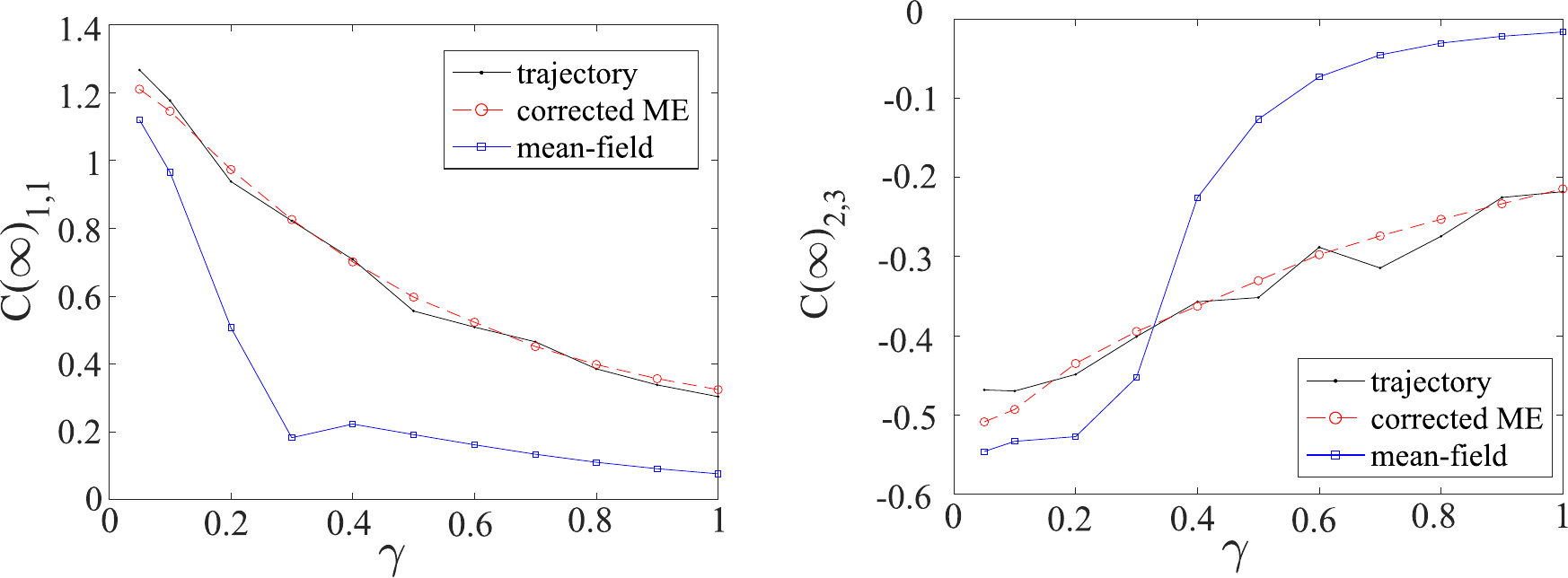}
\caption{Steady state $C$-correlator values as a function of measurement parameter $\gamma$ for a system size of $L=4$  }
\label{fig:C_infinity}
\end{figure*}

Finally, Figure \ref{fig:C_infinity} shows some of the saturated final $C(t=\infty)$ values compared to the trajectory averaged data. The replica cut-off methodology is consistent across all values of the measurement rate.

\section{The replica master equation}
\label{sect:TP}
We now  write the SSE differential equation for the conditional pure-state density matrix as 
\bea
\dd \rho^c_t  &=& -i \dd t [\hat{H}, \rho^c_t]   + \sum_i \dd W_i \left\{ \hat{M}_i, \rho^c_t \right\} -   \\ \non  && \frac{\gamma \dd t}{2} \sum_i  \left\{ \hat{M}^2_i , \rho^c_t \right\} + \sum_{ij} \dd W_i \dd W_j \hat{M}_i \rho^c_t \hat{M}_j
\eea
and then consider the replicated system and compute
\bea
\non
\dd \rho^{R_2}_t &=& \rho^{R_2}_{t+ \dd t} - \rho^{R_2}_t  = \overline{\rho^c_{t+ \dd t} \otimes  \rho^c_{t+ \dd t}} -\overline{\rho^c_{t} \otimes  \rho^c_{t}} \\ &=& \overline{ \dd \rho^c_{t} \otimes  \rho^c_{t}} + \overline{\rho^c_{t } \otimes  \dd \rho^c_{t}}+ \overline{ \dd \rho^c_{t} \otimes \dd \rho^c_{t}}
\eea
we obtain 
\bea
\label{eq:dpall}
\overline{ \dd \rho^c_{t} \otimes  \rho^c_{t}} &=& \dd t \mathcal{L}(\rho_t) \otimes \rho_t \equiv \dd t \mathcal{L}^{(1)} (\rho^{[R_2]}_t) \\ \non
\overline{\rho^c_{t} \otimes \dd \rho^c_{t} } &=& \dd t  \rho_t  \otimes \mathcal{L}(\rho_t) \equiv \dd t \mathcal{L}^{(2)} (\rho^{[R_2]}_t)  \\  \non \overline{ \dd \rho^c_{t} \otimes \dd \rho^c_{t}} &=&  %\overline{ \sum_{ij} \dd W_i  \dd W_j \left\{ \hat{M}^{(1)}_i, \left\{ \hat{M}^{(2)}_j, \rho^c_t \otimes \rho^c_t  \right\} \right\} } \\ && \quad +  \text{ higher order in } \dd t 
 \gamma \dd t \overline {\sum_i \left\{ \hat{M}^{(1)}_i, \left\{ \hat{M}^{(2)}_i, \rho^c_t \otimes \rho^c_t  \right\} \right\} }  \\ && \quad + \text{ higher order in } \dd t \non 
\eea
Expanding this out with
$$
\hat{M}_i = \hat{O}_i + \langle \hat{O}_i \rangle 
$$
gives 
\bea
\label{eq:dpdp}
\overline{ \dd \rho^c_{t} \otimes  \dd \rho^c_{t}} 
% \\ &=& \gamma \dd t \overline {\sum_i \left\{ \hat{O}^{(1)}_i - \langle \hat{O}^{(1)}_i \rangle_t , \left\{ \hat{O}^{(2)}_i - \langle \hat{O}^{2)}_i \rangle_t, \rho^c_t \otimes \rho^c_t  \right\} \right\} }
% \\ \non &=&  \gamma \dd t \sum_i \left\{ \hat{O}^{(1)}_i ,\left\{ \hat{O}^{(2)}_i , \overline{\rho^c_t \otimes \rho^c_t} \right\}\right\}  
%\\  \non && \quad + \gamma \dd t \sum_i 4 \overline{\langle \hat{O}_i^{(1)} \rangle_t \langle \hat{O}^{(2)}_i \rangle_t  \rho^c_t \otimes \rho^c_t} 
%\\ && \non \quad - \gamma \dd t\sum_i 2 \langle \overline{\hat{O}^{(1)}_i \rangle_t \left \{ \hat{O}^{(2)}_i, \rho^c_t \otimes \rho^c_t\right\}} 
% \\ && \non \quad - 2 \langle \overline{\hat{O}^{(2)}_i \rangle_t \left \{ \hat{O}^{(1)}_i, \rho^c_t \otimes \rho^c_t\right\}} \non
 &=&   \gamma \dd t\sum_i \left\{ \hat{O}^{(1)}_i ,\left\{ \hat{O}^{(2)}_i , \overline{\rho^c_t \otimes \rho^c_t} \right\}\right\}   \\ && \quad - 2 \gamma \dd t \sum_i  \left \{ \hat{O}^{(1)}_i+\hat{O}^{(2)}_i,\overline{ \langle \hat{O}_i \rangle_t  \rho^c_t \otimes \rho^c_t} \right\} \non \\ && \non \quad + 4 \gamma \dd t \sum_i  \overline{\langle \hat{O}_i \rangle_t^2  \rho^c_t \otimes \rho^c_t} 
\eea
where $\hat{O}_i^{(1)}= \hat{O}_i \otimes I$,  $\hat{O}_i^{(2)}= I\otimes \hat{O}_i  $ are operators acting on different replica subspaces and $ \langle \hat{O}_i \rangle_t  \equiv \langle \hat{O}^{(1)} _i \rangle_t =\langle \hat{O}^{(2)} _i \rangle_t$ because of the inherent symmetry between replicas. 

\subsection{An infinite hierarchy and the need for a cutoff}

The latter two terms cause a problem because it is not possible to fully disentangle statistical correlations between the expectation values $\langle \hat{O} _i \rangle$ and $\rho^c_t \otimes \rho^c_t$.  However, it was pointed out  \cite{Buchhold2021}  that these terms could be calculated by using higher-order replicas:
$$
\overline{ \langle \hat{O}_i \rangle_t  \rho^c_t \otimes \rho^c_t}  = \Tr_{(3)} \left[ \hat{O}_i^{(3)} \rho_t^{R_3} \right]
$$
and
$$
\overline{ \langle \hat{O}_i \rangle_t^2  \rho^c_t \otimes \rho^c_t}  = \Tr_{(3,4)} \left[ \hat{O}_i^{(3)} \hat{O}_i^{(4)} \rho_t^{R_4} \right]
$$
whereby $\rho_t^{R_3}$ and $\rho_t^{R_4}$ are the three and four replica density matrices. An apparent problem arises in that to calculate the 2-replica, you need to also calculate the 3 and 4 replicas - and to calculate them, you need higher-order replicas and so on - setting off an infinite hierarchy of replica dependency. 

An interesting and practical workaround is to use the notion of a mean-field decoupling  \cite{Buchhold2021} such that we replace these terms with
\bea
\overline{ \langle \hat{O}_i \rangle_t \rho^c_t \otimes \rho^c_t } &\rightarrow&  \overline{ \langle \hat{O}_i \rangle_t } \times  \overline{  \rho^{c\phantom |}_t \otimes \rho^{c\phantom |}_t }  = \overline{ \langle \hat{O}_i \rangle_t } \times  \rho^{R_2}  \non  \\  \overline{ \langle \hat{O}_i \rangle_t^2 \rho^c_t \otimes \rho^c_t } &\rightarrow & \overline{ \langle \hat{O}^{(1)}_i \hat{O}^{(2)}_i  \rangle_t } \times  \overline{  \rho^{c\phantom |}_t \otimes \rho^{c\phantom |}_t }  \\ && \quad \non  =\overline{ \langle \hat{O}^{(1)}_i \hat{O}^{(2)}_i  \rangle_t } \times  \rho^{R_2}    \non
\eea
The final expression that is 
\bea
\label{eq:dpR2}
\dd \rho^{R_2}_t  &=& \non  (\mathcal{L}^{(1)} + \mathcal{L}^{(2)} - 4 \gamma \overline{C_t} ) \dd t  \rho_t^{R_2}     \\ &+ & \dd t \gamma \sum_i \left\{ \hat{O}^{(2)}_i - \overline{\langle \hat{O}^{(2)}_i\rangle} ,\left\{  \hat{O}^{(1)}_i - \overline{\langle \hat{O}^{(1)}_i\rangle}, \rho^{R_2}_t]\right\} \right\} \non
\eea
where 
  $$\overline{C_t} = \sum_i \overline{ \langle \hat{O}^{(1)}_i \hat{O}^{(2)}_i \rangle_t} - \overline{\langle \hat{O}^{(1)}_i \rangle_t } \;\;  \overline{\langle \hat{O}^{(2)}_i \rangle_t }. $$
Although this may be well motivated for many scenarios,  a key problem with this shown next is that it does not preserve partial traces.  %The motivation for this note is to show that one can calculate these quantities can be estimated accurately from knowledge of the 2-replica alone - allowing us to cut off the replica expansions right at the start. 

\subsection{Trace and partial trace preservation}

The term $4 \gamma \overline{C}_t$ in the mean-field replica cutoff is included above to make the update trace-preserving, and directly cancels the three terms:
\bea
&& \non \Trace  \sum_i \left\{ \hat{O}_i^{(1)} ,\left\{ \hat{O}_i^{(2)} , \rho^R_t  \right\}   \right\} =4  \sum \overline{  \langle \hat{O}_i^{(1)} \hat{O}_i^{(2)} } \rangle_t \\ 
&& \Trace   \left[ -2 \sum_i  \left \{ \hat{O}^{(1)}_i+\hat{O}^{(2)}_i,\overline{ \langle \hat{O}_i \rangle_t }  \rho^{R_2}_t \right\} \right] \\ && \quad =  - 8 \sum_i \overline{\langle \hat{O}_i^{(1)}  \rangle_t} \;\; \overline{\langle \hat{O}_i^{(2)}  \rangle_t} \non \\
&& \non \Trace \sum_i 4 \overline{\langle \hat{O}_i^{(1)} \rangle_t \langle \hat{O}^{(2)}_i \rangle_t } \rho^{R_2}_t  =  4 \sum_i \overline{\langle \hat{O}_i^{(1)}  \rangle_t} \;\; \overline{\langle \hat{O}_i^{(2)}  \rangle_t}  \non
\eea
However, the mean-field decoupling does not preserve the partial trace with respect to either of the replica copies. If we define $\hat{\chi}_i = \Trace_{(k)} \left( \hat{O}_i^{(k)} \rho^{R_2} \right)$  where $\Trace_{(k)} $ is the partial trace over the $\text{k}^\text{th}$ replica (in this case $k=1$ or $k=2$) we see that 
\bea
 && \Trace_{(k)}  \sum_i \left\{ \hat{O}_i^{(1)} ,\left\{ \hat{O}_i^{(2)} , \rho^R_t  \right\}   \right\}  = 2  \sum_i \left\{  \hat{O}_i , \hat{\chi}_i \right\} \non  \\ 
&& \Trace_k   \left[ -2 \sum_i  \left \{ \hat{O}^{(1)}_i+\hat{O}^{(2)}_i,\overline{ \langle \hat{O}_i \rangle_t }  \rho^{R_2}_t \right\} \right] \\&& \quad = -2 \langle \hat{O}_i \rangle \left( \left\{ \hat{O}_i ,\rho_t \right\} + 2 \hat{\chi}_i \right) \non \\
&& \Trace_k \sum_i 4 \overline{\langle \hat{O}_i^{(1)} \rangle_t \langle \hat{O}^{(2)}_i \rangle_t } \rho^{R_2}_t  =  4 \sum_i \langle \hat{O}_i \rangle^2_t \rho_t \non
\eea
and where we use $\Tr_k \rho^{R_2}_t = \rho_t$ throughout.  These terms do not generally cancel each other out, so the mean-field cutoff does not faithfully preserve the dynamics of a single copy. 
  
Contrast this with the behaviour of the original terms in \eqref{eq:dpdp} where we can see that tracing out either copy results in terms that end up cancelling each other. Suppose for simplicity we trace out the second replica. We get 
\bea
\non
&& \Trace_{(2)} \overline{ \dd \rho^c_{t} \otimes  \dd \rho^c_{t}}  = 2 \gamma \dd t  \sum_i \left\{ \hat{O}_i, \chi_i \right\}   \\ &&  \non - 2 \gamma \dd t  \sum_i  \left\{ \hat{O}_i, \chi_i \right\}  - 4 \gamma \dd t   \sum_i \Trace_{(2,3)} ( \hat{O}^{(2)}_i \hat{O}^{(3)}_i  \rho^{R_3} ) \\  &&
 + 4 \gamma \dd t  \sum_i \Trace_{(2,3,4)} ( \hat{O}^{(3)}_i \hat{O}^{(4)}_i  \rho^{R_4} ) =0
\eea
where we use the permutation and reduction properties of the replica states to cancel the last two terms.  Taking the partial trace of the other two terms in \eqref{eq:dpall} we see that one of them will always vanish to leave the usual GKSL dynamics on a single replica. 

The key idea behind our approach is to use the permutation symmetry of the replica and the fact that expectation values calculated using $\rho^{R_2}$ must also be valid for higher order replicas $\rho^{R_3}$  and $\rho^{R_4}$.  This allows us to exactly project information at the $R_2$ level to the higher order replicas. Moreover, we will see that it is also possible to positivity requirement to make better estimates of $ {\rho}^{R_N}$  and thus in turn, more accurately model the quantities 
\bea
\label{eq:TracerhoE}
\overline{ \langle \hat{O}_i \rangle_t  \rho^{(c)} \otimes \rho^{(c)}}  &\approx& \Tr_{(j)} \left[ \hat{O}_i^{(j)} \rho_{E}^{R_3} \right] \\
\overline{ \langle \hat{O}_i \rangle_t^2  \rho^{(c)} \otimes \rho^{(c)}_t}  &\approx& \Tr_{(j,k)} \left[ \hat{O}_i^{(j)} \hat{O}_i^{(k)} \rho^{R_4}_{E} \right] \non
\eea
without necessarily needing to model the dynamics of the higher-order replicas directly.

\section{Projected overlap method}
\label{sect:projected_overlap_method}.
The problem of partial trace preservation does not arise if one uses the original expressions \eqref{eq:T3T4}.  However, the requirement that the higher replica states be exact is overly restrictive. Indeed, there is the freedom to choose arbitrary modification of the higher $\rho^{R_n}$ provided they are null under the partial trace reduction to the next lowest replica state $\rho^{R_{n-1}}$.  

One of the most direct is to simply represent the replica density matrix in terms of its Hilbert-Schmidt projections of possible observables.  We can consider, for example, the Pauli group that we can construct from single Pauli-operators $\{ X,Y,Z, I \}$operating on our $L \times M$ qubits of the $M$-replica - we label these generally as $\hat{O}_j$.  Another option would be to consider the canonical representation $\Gamma_j$ of all possible combinations of Majorana operators. 
In what follows, we will work with a vectorised notion of the Hilbert-Schmidt inner product.
$$
\left(A | B\right) \equiv\Tr A^\dagger  B . %{  \sqrt{ \Tr A^\dagger A \Tr B^\dagger B } } =  \delta_{A,B}
$$

Using this notation, if  we have access to all of the M-replica Hilbert space, then we can write our density matrix in the orthogonal operator basis as
\bea
| \rho^{R_M} ) &=&
 \frac{1}{4^{L M}} \sum_{j=1}^{4^{LM}}  (\hat{O}_j  | \rho^{R_M} ) |  \hat{O}_j ) % \\ &=&  \sum_{j=1}^{4^{LM}} \langle O_j \rangle \hat{O}_j  
\eea
where the index $j$ runs over all $4^{LM}$ orthogonal operators.   However, suppose we can only access a subset of the replica spaces $N<M $. Then, we cannot fully know all the weights we need to represent an exact version of the state. In this case, an estimate of the replica could be made for example, by only including the weights that we can know for sure
\bea
|\rho_{E}^{R_M}) &=&  \frac{1}{4^{L M}} \sum_{j=1}^{S_N}  (\hat{O}_j  | \rho^{R_M} ) |  \hat{O}_j ) %\Trace (\hat{O}_j  \times \rho^{[R_N]} )   \hat{O}_j \\  &=& \frac{1}{2^{L M}} \sum_{j=1}^{S_N}  \langle O_j \rangle \hat{O}_j
\eea
where $S_N$ is the number of operators out of the $4^{LM}$ that act nontrivially on a maximum of $N<M$ replicas.  As we will discuss, we may also consider other factors in an attempt to fill in some of the missing information. One such situation that arises is that the  $\rho_E^{R_M} $ as given above may not be properly physical and have negative eigenvalues. This allows us to try to make estimates of the parameters that we don't know, improving our guess of the state even further.

\subsection{Projection to permutation symmetric sub-spaces}

We want to embed the above idea in a formalism that reflects the underlying permutation symmetry of the replicas and any symmetry of the bare system (e.g. number conservation). This permutation basis is analogous to the Bloch-basis methods - but where instead of just translation symmetry and zero total momentum, we must account for the fact that the replicated spaces are invariant under replica permutation.  

The $i^{th}$ basis vector of these states is denoted $\ket*{V^{R_N}_i}$ and takes its dimensionality from the full Hilbert space of the N-replica  $2^{(LN)}$ where $L$ is the number of the qubits in a single copy and $N$ is the number of copies.  The number of unique vectors needed depends on the number of qubits in the bare system and whether there are any additional symmetries.  For two qubits with number conservation at half-filling, and sub-lattice symmetry, we have $D_N=\text{dim-} V^{[R_N]}$ with  $D_1= 2, D_2= 3, D_3= 4, D_4 =  5 $. Similarly, for four qubits with half filling, we have  $D_1= 6, D_2 = 21, D_3=56, D_4 = 126$.
%\nin $ \text{dim-} V^{[R_1]} = (2^2,2) , \text{dim-} V^{[R_2]} = (2^4, 3), \text{dim-} V^{[R_3]} = (2^6 , 4), \text{dim-} V^{[R_4]} = (2^8,  5) $. Similarly, for four qubits with half filling, we have
%\nin $ \text{dim-} V^{[R_1]} = (4^2,6), \text{dim-} V^{[R_2]} = (4^4, 21), \text{dim-} V^{[R_3]} = (4^6,  56), \text{dim-} V^{[R_4]} =  (4^8, 126)$. 

The density matrices $\rho^{R_N}$ can be projected fully to the corresponding symmetric sub-space 
\bea
\rho^{R_N} = P^{R_N} \rho^{R_N}  P^{R_N} .
\eea
where $P^{R_N} = \sum_i |V^{R_N}_i \rangle \langle {V^{R_N}_i} | $
and we define the projected matrix representation as
\bea
r^{R_N}_{ij} = \bra*{V^{R_N}_i}  \rho^{R_N}   \ket*{V^{R_N}_j}
\eea

It is also useful to know part how physical observables project to the symmetric replica spaces. 
For the orthonormal set of operators $\hat{O}$, we define the projected operators as
\bea
\hat{o}^{R_N}_{ij} = \bra*{V^{R_N}_i}  \hat{O}  \ket*{V^{R_N}_j}
\eea
Although the original operators $\hat{O}_i$ form an orthonormal set
\be
( \hat{O}_i | \hat{O}_j)  \equiv \frac{1}{2^{LN}} \Trace ( \hat{O}_i^\dagger \hat{O_j}) = \delta_{i,j}
\ee
the projected operators $\hat{o}^{R_N}$ do not.  We will denote with $\hat{O}^{V_N}$ a  $S_N$ dimensional subset of the $\hat{O}$ operators that, when projected to $\ket{V^{R_N}}$, are independent in a Hilbert-Schmidt sense.   We then denote these linearly independent, subspace-projected operators as $\hat{o}^{V_N}$. 
$$
\hat{o}_i^{V_N} = \bra*{V^{R_N}}  \hat{O}_i^{V_N}  \ket*{V^{R_N}}
$$
where the $\hat{O}_i^{V_N}$ are a subset of the original operators $\hat{O}_i$. 

As a general rule, this set $\hat{O}^{V_N}$ will also be linearly independent when projected to a higher replica symmetric space  (e.g. $\ket*{V^{R_{M}}}$ with $M>N$).   This allows us to successively build up larger sets of independent operators from those that are independent on lower-order replicas:
$$
\hat{O}^{V_M} = \{ \hat{O}^{V_{M-1} }, \text{ + operators existing only on }  R_M \}
$$
 
In the vectorised representation $\hat{o}^{V_N} \rightarrow | o^{V_N} )$ we can write now $| \tilde{o}_i^{V_N})$ as the orthonormalised set $| o_i^{V_N})$ such that 
$(  \tilde{o}^{V_N}_i | \tilde{o}^{V_N}_j) = \delta_{ij}$. Writing $ C^N_{ij}= ( o^{V_N}_i | \tilde{o}^{V_N}_j)$ as the overlap matrix between these non-orthogonal and orthonormal spaces, we then relate the expectation values of the operators $\hat{O}^{V_N}$ to expectation values of $ \hat{\tilde{o}}^{V_N}$ via:
$$
( {\tilde{o}}^{V_N}_i | r^{R_N} ) =  \sum_j [C^N]^{-1}_{i,j} ( \hat{O}^{V_N}_j |  \rho^{R_N})
$$
allowing one to relate full replica space expectation values $( \hat{O}^{V_N}_j |  \rho^{R_N})$ to those of the orthonormal symmetric subspace.  This then allows us to write 
\bea
\non
|r^{R_N})  &=& \sum_i  ( \tilde{o}_i^{V_N}| r^{R_N} ) \times | \tilde{o}_i^{V_N} )  =  \sum_i  r^{R_N} (\tilde{o}_i) \times | \tilde{o}_i^{V_N} )  \\
 &=& \sum_{ij}  [C^N]^{-1}_{i,j} ( \hat{O}^{V_N}_j |  \rho^{R_N})  \times  | \tilde{o}_i^{V_N} ) \non 
\eea
giving us a direct way to relate the projected replica density matrix to expectation values calculated in the unprojected space. Indeed, assuming as described above, we use the projected linear independent operators from lower replicas as part of the higher replica independent sets, we can use a decomposition of $r^{R_{N}}$ as above to partially construct the $r^{R_{M}}$ (with $M>N$) such that the correct partial trace information is properly encoded higher replica level.  
\be
\non
( \tilde{o}_i^{V_M}| r^{R_M}_E ) =  \sum_{jk}[C_{S_N}^M]^{-1}_{i,j} [C^N]_{j,k} ( \tilde{o}_k^{V_N}| r^{R_N} )
\ee
where $C_{S_N}^M$ is the matrix containing the first $S_N$ rows and columns of $ C^M_{ij}= ( o^{V_M}_i | \tilde{o}^{V_M}_j)$

This construction guarantees that the $N$-replica information is correctly transposed onto the $M>N$ replica estimate.  Returning to a non-vectorised picture, we write
\be
r_E^{R_M} = \sum_i ( \tilde{o}_i^{V_M}| r^{R_M}_E ) \times \tilde{o}_i^{V_M}
\ee
and then 
\bea
\label{eq:rhoV}
\rho^{R_{M}}_{E}  &=& \ket*{V^{R_{M}}} r_{E}^{R_{M}}  \bra*{V^{R_{M}}} 
\\ &\equiv &  \sum_{i,j}   [r_{E}^{R_{M}}]_{i,j} \ket*{V^{R_{M}}_i}  \bra*{V^{R_{M}}_j} .
\eea
This construction guarantees the property
\be
\Tr_{(a,b)}  \rho^{R_{N+2}}_{E}  = \Tr_{(a)}  \rho^{R_{N+1}}_{E}  = \rho^{R_{N}} .
\ee

\section{Null spaces and eigenbases}
 \label{sect:eigenbasis_method}
Here, we outline an alternative way to construct null spaces of partial traces in the replica setup, specifically showing how the eigenstructure of $\rho^{R_n}$ can be used to assemble the necessary structure at higher order replicas such that they reduce to $\rho^{R_n}$ under partial trace. We also discuss how to navigate the partial trace null spaces of higher replica states within this alternative framework.  

Using this approach, to approximate $\rho^{R_3}$ (and $\rho^{R_4}$) using only information from $\rho^{R_2}$ we first 
 perform an eigendecomposition of $\rho^{R_2}$:
   
\begin{equation}
\rho^{R_2} = \sum_n p_n \ket{v_n}\bra{v_n}
\end{equation}
    
Due to the permutation of the replicated $\rho^{R_2}$ for each eigenvector $\ket{v_n}$ we can perform a Schmidt decomposition such that
\begin{equation}
\ket{v_n} = \sum_{j=1}^{N_s} s^n_j \ket{A_j}\ket{A_j} =\sum_{j=1}^{N_s} s^n_j \ket{A^n_j A^n_j}
\end{equation}
 where $\sum_j (s^n_j)^2 =1$.  
    
The reconstructed $\rho^{R_2}$ can be represented by the following triple sum:
\bea
&& \rho^{R_2}=  \sum_{n=1}^{N_p} p_n  \times \sum_{j=1}^{N_s} s^{n}_j \ket{A^n_j  A^n_j}  \times  \sum_{k=1}^{N_s} s^{n*}_k \bra{A^n_k A^n_k}\non
\eea
where $N_p$ is the number of eigenvalues, $N_s$ is the number of terms in the Schmidt decomposition, $p_i$ are the eigenvalues, and the real $s^n_j$ are the Schmidt coefficients for the $n$-th eigenvector.

The sum can be divided into incoherent  and coherent sums:
$$
\rho^{R_2} = \rho^{R_2}_{\text{I}}  + \rho^{R_2}_{\text{C}} 
$$
with
$$
 \rho_{\text{I}} = \sum_{n=1}^{N_p} p_n \sum_{j=1}^{N_s}  (s^n_j )^2  \ket{A^n_j A^n_j} \bra{ A^n_j A^n_j}
$$
and 
$$
\rho_{\text{C}} = \sum_{n=1}^{N_p} \sum_{j=1}^{N_s} \sum_{k=j+1}^{N_s}  p_n  s^n_j  s^{n}_k  ( \ket{A^n_j A^n_j} \bra{ A^n_k A^n_k} + h.c.)
$$
For notational purposes, we now write $\ket{A}=\ket{A_j}$ and $\ket{B}= \ket{A_k} $ and observe that the coherent sum runs over pairs of states in the form $\ket{AA} \bra{BB}$ whereby construction $\braket{A}{B} =0 $. 

Suppose we want now to construct higher replicated density matrices such that they give the correct incoherent and coherent terms. To encode the permutation symmetry, we consider states of the form  
\bea 
&& \ket{3,0} \equiv \ket{AAA}  \non \\ 
&& \ket{2,1} \equiv  \mathring{\ket{AAB}} = \ket{AAB}+\ket{ABA}+\ket{BAA} \non \\
&& \ket{2,1} \equiv  \mathring{\ket{BBA}} = \ket{BBA}+\ket{BAB}+\ket{ABB} \non \\
&&  \ket{0,3} \equiv \ket{BBB} \non
\eea
for $R_{3}$ and 
\bea 
&& \ket{4,0} \equiv \ket{AAAB} \non \\ 
&& \ket{3,1} \equiv  \mathring{\ket{AAAB}} = \ket{AAAB}+ \text{permutations} \non \\
&& \ket{2,2} \equiv  \mathring{\ket{AABB}} = \ket{AABB}+ \text{permutations} \non \\
&&  \ket{1,3} \equiv \mathring{\ket{BBBA}} = \ket{BBBA}+ \text{permutations} \non \\
&& \ket{0,4} \equiv \ket{BBBB} \non \\ 
\eea
for $R_{4}$ where we deliberately exclude normalisation factors that we might be tempted to add.

Focusing on the behaviour of $R_3$ under partial trace to $R_2$ we see, for example, that 
$$ \Tr_{1} \op{AAA}{AAA} = \op{AA}{AA}.$$
We therefore have a natural way to construct $R^3$ analogues of the incoherent $\rho^{R_2}_I$. That is 
\be
\rho^{R_2}_I =  \Tr_{1} \left[   \sum_{n=1}^{N_p} p_n \sum_{j=1}^{N_s}  (s^n_j )^2  \ket{A^n_j A^n_j A^n_j} \bra{ A^n_j A^n_j A^n_j }  \right] \non
\ee
Likewise, to construct coherent terms, we can observe that
$$ 
\Tr_{1} \ket{AAA} \mathring{\bra{BBA}} =  \Tr_1 \mathring{\ket{AAB}}\bra{BBB} = \op{AA}{BB}  $$
We could then, for example, write 
\begin{align}
\non \rho^{R_2}_C &=  \Tr_{1}  \left[  \sum_{n=1}^{N_p} \sum_{j=1}^{N_s} \sum_{k=j+1}^{N_s}  p_n  s^n_j  s^{n}_k  \right.  \\   \quad & ( (\frac{1}{2} + \alpha) \ket{A^n_j A^n_j A^n_j } \mathring{\bra{ A^n_j A^n_k A^n_k }} \\ & \quad  \left.  \phantom\sum^{\phantom f} +  (\frac{1}{2} - \alpha)  \mathring{\ket{A^n_j A^n_j A^n_k } }\bra{ A^n_k A^n_k A^n_k } + h.c.) \right] \non 
\end{align}
The freedom here to choose $\alpha$ and still retain the same state under partial trace illustrates a general feature where there are operators in the larger replicated space that are null under the partial trace operation.  

To find these null operator spaces, we use an alternative representation that uses bosonic notation, where creation operators $a^\dagger$ and $b^\dagger$ correspond to states A and B, respectively. 
States are represented as $\ket{n_a, n_b}$, where $n_a$ and $n_b$ are the number of a and b bosons. For example, $\ket{3,1}$ is equivalent to $(a^\dagger)^3b^\dagger\ket{0}$, corresponding to the symmetrised state of three A's and a single B. The partial trace process on $A/B$ subspaces can be represented using these operators:
\bea
\non \Tr_{N}  \left[ \boldsymbol{\cdot} \right]  &=& a^N \boldsymbol{\cdot} (a^\dagger)^N + b^N \boldsymbol{\cdot} (b^\dagger)^N
\eea
where $N$ is the number of replica subspaces we are tracing out. The previous calculation  
$$ \Tr_{1} \ket{AAA} \mathring{\bra{BBA}} = \op{AA}{BB}  $$ becomes particularly simple in this notation
\bea  \Tr_{1} [ \op{3,0}{1,2} ] &=& a \op{3,0}{1,2} a^\dagger  +  b \op{3,0}{1,2} b^\dagger \non   \\ &=& \op{2,0}{0,2}  \non 
\eea

\subsection{Constructing higher replica null spaces}
We can now apply these operations to all outer product combinations in our symmetric $R_3$ and $R_4$ vector spaces.  These results are catalogued in the tables below.  In TABLE \ref{tab:R3_to_R2_partial_trace}, we show how $R_3$ behaves under reduction to $R_2$.  The crucial observation to make in this table (and the others) is the relationship between elements along the main diagonal and those off-diagonals parallel to it.  In particular, note that every entry has a repetition in the diagonally $\nwarrow \searrow$ adjacent elements. From here, it is easy to see that summing along the diagonals with alternating signs will give an operator in $R^3$ that vanishes when we trace out any one of the replica subspaces.  The full set of Hermitian null space operators when going from R3 to R2 is there:
\begin{align*}
N_{3\to2}^{(1)} &= \ket{3,0}\bra{3,0} - \ket{2,1}\bra{2,1} + \ket{1,2}\bra{1,2} - \ket{0,3}\bra{0,3} \\
N_{3\to2}^{(2)} &= \ket{3,0}\bra{2,1} - \ket{2,1}\bra{1,2} + \ket{1,2}\bra{0,3} +h.c. \\
N_{3\to2}^{(3)} &= \ket{3,0}\bra{1,2} - \ket{1,2}\bra{0,3} +h.c. \\
N_{3\to2}^{(4)} &= \ket{3,0}\bra{0,3} +h.c. 
\end{align*}
where  $N_{3\to2}^{(3)}$ is the space associated with the obseved freedom to choose $\alpha$ above. 

In  TABLE \ref{tab:R4_to_R3_partial_trace}, we show the analogue construction for $R^4 \rightarrow R^3$, and we can observe the same pattern on the diagonals. The resulting  null spaces are
\begin{align*}
N_{4\to3}^{(1)} &= \ket{4,0}\bra{4,0} - \ket{3,1}\bra{3,1} + \ket{2,2}\bra{2,2}  ...  \\ & \quad \quad \quad \quad - \ket{1,3}\bra{1,3} + \ket{0,4}\bra{0,4} \\
N_{4\to3}^{(2)} &= \ket{4,0}\bra{3,1} - \ket{3,1}\bra{2,2} + \ket{2,2}\bra{1,3} ... \\ & \quad \quad\quad \quad - \ket{1,3}\bra{0,4} + h.c. \\
N_{4\to3}^{(3)} &= \ket{4,0}\bra{2,2} - \ket{3,1}\bra{1,3} + \ket{2,2}\bra{0,4}  +h.c. \\
N_{4\to3}^{(4)} &= \ket{4,0}\bra{1,3} - \ket{3,1}\bra{0,4}  +h.c. \\
N_{4\to3}^{(5)} &= \ket{4,0}\bra{0,4} + h.c. 
\end{align*}

Finally, we note that the $R_3 \rightarrow R_2$ null space operators also have a representation within $R_4$.  These can be calculated as follows
\begin{align*}
N_{3\to2}^{(1)} &= \ket{3,1}\bra{3,1} - \ket{1,3}\bra{1,3} ... \\ & \quad \quad - 2(\ket{4,0}\bra{4,0} + \ket{0,4}\bra{0,4}) \\
N_{3\to2}^{(2)} &= \ket{3,1}\bra{2,2} - \ket{2,2}\bra{1,3} - 3(\ket{4,0}\bra{3,1}  ... \\ & \quad \quad +\ket{1,3}\bra{0,4}) + h.c. \\
N_{3\to2}^{(3)} &= \ket{4,0}\bra{2,2} - \ket{2,2}\bra{0,4} + h.c. \\
N_{3\to2}^{(4)} &= \ket{4,0}\bra{1,3} + h.c.
\end{align*}
from which we obtain the $N_{3\to2}$ expressions above under a single partial trace.

\subsubsection{A discussion on the $R_2 \rightarrow R_1$ problem}

Assuming we can calculate $\rho^{R_2}$ we can write
$$
 \rho^{R_2}_{\text{I}} = \sum_{n=1}^{N_p} p_n \sum_{j=1}^{N_s}  (s^n_j )^2  \ket{A^n_j A^n_j} \bra{ A^n_j A^n_j}
$$
and 
$$
\rho^{R_2}_{\text{C}} = \sum_{n=1}^{N_p} \sum_{j=1}^{N_s} \sum_{k=j+1}^{N_s}  p_n  s^n_j  s^{n}_k  ( \ket{A^n_j A^n_j} \bra{ A^n_k A^n_k} + h.c.)
$$
On partial trace reduction from $R^2$ to $R^1$, we find that only the diagonal incoherent term survives because  
\bea
\Tr_1 \op{AA}{BB} = \Tr_1 \op{BB}{AA} =   0  
\eea
or in the other notation
\bea
\Tr_1 \op{2,0}{0,2} = \Tr_1 \op{0,2}{0,2} =   0  
\eea

For the incoherent parts, it is a bit redundant to speak of A's and B's, but we could write it as
\bea
\Tr_1 \op{AA}{AA} =   \op{A}{A}  
\eea
or in the other notation
\bea
\Tr_1 \op{2,0}{2,0} = \op{1,0}{1,0}   
\eea
 such that
 $$
\rho^{R_1}= \Tr_1\left[ \rho^{R_2}_{\text{I}}\right]  = \sum_{n=1}^{N_p}  \sum_{j=1}^{N_s}  p_n \times (s^n_j )^2  \ket{A^n_j } \bra{ A^n_j }
$$
and where it is important to note that in general  $\braket{A^n_j }{A^m_k} \ne 0$ for $n \ne m$ .  

For each pair $(A,B)\equiv (A^n_j, A^n_k)$, $j \ne k$  are three operators that we can construct that vanish under the partial trace of one of the two subsystems, see table \ref{tab:R2_to_R1_partial_trace}.
\begin{align*}
N_{2\to1}^{(1)} &= \ket{2,0}\bra{2,0} - \ket{1,1}\bra{1,1}  + \ket{0,2}\bra{0,2}  \\
N_{2\to1}^{(2)} &= \ket{2,0}\bra{1,1} - \ket{1,1}\bra{0,2} +h.c. \\
N_{2\to1}^{(3)} &= \ket{2,0}\bra{0,2} + h.c. \\
\end{align*}
The last one is of the form $ \op{AA}{BB}$, which by construction forms the coherent part of $\rho^{R_2}$. What about the other two? They don't play any role in the construction of $\rho_{R_2}$ using this method. In the other notation, they are off the form 
\begin{align*}
N_{2\to1}^{(1)} &= \ket{AA}\bra{AA} - \mathring{\ket{AB}}\mathring{\bra{AB}}  + \ket{BB}\bra{BB}  \\
N_{2\to1}^{(2)} &= \ket{AA}\bra{AB} - \ket{AB}\bra{BB} +h.c. \\
\end{align*}
The common feature here are terms containing states like $\mathring{\ket{AB}}$, which are not obtained with the $R_2$ decomposition. 

For completeness, let's examine the $R^3$ higher replica analogue of this null space.
\begin{align*}
N_{3\to1}^{(1)} &= \frac{1}{2} (3\ket{3,0}\bra{3,0} - \ket{2,1}\bra{2,1}  \\ & \quad \quad - \ket{1,2}\bra{1,2} + 3 \ket{0,3} \bra{0,3} )  \\
N_{3\to1}^{(2)} &= \ket{3,0}\bra{2,1} - \ket{1,2}\bra{0,3} +h.c. \\
N_{3\to1}^{(3)} &= \ket{3,0}\bra{1,2} - \ket{3,0}\bra{1,2} +h.c. \\
\end{align*}

\begin{table*}
\centering
\begin{tabular}{!{\vrule width 1pt}c!{\vrule width 1pt}c!{\vrule width 1pt}c!{\vrule width 1pt}c!{\vrule width 1pt}}
\noalign{\hrule height 1pt}
& $\bra{2,0}$ & $\bra{1,1}$ & $\bra{0,2}$ \\
\noalign{\hrule height 1pt}
$\ket{2,0}$ & $\op{1,0}{1,0}$ & $\op{1,0}{0,1}$ & $0$ \\
\noalign{\hrule height 1pt}
$\ket{1,1}$ & $\op{0,1}{1,0}$ & $\op{0,1}{0,1} + \op{1,0}{1,0}$ & $\op{1,0}{0,1}$ \\
\noalign{\hrule height 1pt}
$\ket{0,2}$ & $0$ & $\op{0,1}{1,0}$ & $\op{0,1}{0,1}$ \\
\noalign{\hrule height 1pt}
\end{tabular}
\caption{Transformation of R2 basis states under partial trace to R1 states}
\label{tab:R2_to_R1_partial_trace}
\end{table*}

\begin{table*} 
\centering
\begin{tabular}{!{\vrule width 1pt}c!{\vrule width 1pt}c!{\vrule width 1pt}c!{\vrule width 1pt}c!{\vrule width 1pt}c!{\vrule width 1pt}}
\noalign{\hrule height 1pt}
 & $\bra{3,0}$ & $\bra{2,1}$ & $\bra{1,2}$ & $\bra{0,3}$ \\
\noalign{\hrule height 1pt}
$\ket{3,0}$ & $\op{2,0}{2,0}$ & $\op{2,0}{1,1}$ & $\op{2,0}{0,2}$ & $0$ \\
\noalign{\hrule height 1pt}
$\ket{2,1}$ & $\op{1,1}{2,0}$ & $\op{1,1}{1,1} + \op{2,0}{2,0}$ & $\op{1,1}{0,2} + \op{2,0}{1,1}$ & $\op{2,0}{0,2}$ \\
\noalign{\hrule height 1pt}
$\ket{1,2}$ & $\op{0,2}{2,0}$ & $\op{0,2}{1,1} + \op{1,1}{2,0}$ & $\op{0,2}{0,2} + \op{1,1}{1,1}$ & $\op{1,1}{0,2}$ \\
\noalign{\hrule height 1pt}
$\ket{0,3}$ & $0$ & $\op{0,2}{2,0}$ & $\op{0,2}{1,1}$ & $\op{0,2}{0,2}$ \\
\noalign{\hrule height 1pt}
\end{tabular}
\caption{Transformation of $R^3$  symmetric states to $R^2$ under single partial trace}
\label{tab:R3_to_R2_partial_trace}
\end{table*}

\begin{table*}
\centering
\begin{tabular}{!{\vrule width 1pt}c!{\vrule width 1pt}c!{\vrule width 1pt}c!{\vrule width 1pt}c!{\vrule width 1pt}c!{\vrule width 1pt}c!{\vrule width 1pt}}
\noalign{\hrule height 1pt}
 & $\bra{4,0}$ & $\bra{3,1}$ & $\bra{2,2}$ & $\bra{1,3}$ & $\bra{0,4}$ \\
\noalign{\hrule height 1pt}
$\ket{4,0}$ & $\op{3,0}{3,0}$ & $\op{3,0}{2,1}$ & $\op{3,0}{1,2}$ & $\op{3,0}{0,3}$ & $0$ \\
\noalign{\hrule height 1pt}
$\ket{3,1}$ & $\op{2,1}{3,0}$ & $\op{2,1}{2,1} + \op{3,0}{3,0}$ & $\op{2,1}{1,2} + \op{3,0}{2,1}$ & $\op{2,1}{0,3} + \op{3,0}{1,2}$ & $\op{3,0}{0,3}$ \\
\noalign{\hrule height 1pt}
$\ket{2,2}$ & $\op{1,2}{3,0}$ & $\op{1,2}{2,1} + \op{2,1}{3,0}$ & $\op{1,2}{1,2} + \op{2,1}{2,1}$ & $\op{1,2}{0,3} + \op{2,1}{1,2}$ & $\op{2,1}{0,3}$ \\
\noalign{\hrule height 1pt}
$\ket{1,3}$ & $\op{0,3}{3,0}$ & $\op{0,3}{2,1} + \op{1,2}{3,0}$ & $\op{0,3}{1,2} + \op{1,2}{2,1}$ & $\op{0,3}{0,3} + \op{1,2}{1,2}$ & $\op{1,2}{0,3}$ \\
\noalign{\hrule height 1pt}
$\ket{0,4}$ & $0$ & $\op{0,3}{3,0}$ & $\op{0,3}{2,1}$ & $\op{0,3}{1,2}$ & $\op{0,3}{0,3}$ \\
\noalign{\hrule height 1pt}
\end{tabular}
\caption{Transformation of $R^4$  symmetric states to $R^3$ under single partial trace}
\label{tab:R4_to_R3_partial_trace} 
\end{table*}

\begin{table*}
\centering
\begin{tabular}{!{\vrule width 1pt}c!{\vrule width 1pt}c!{\vrule width 1pt}c!{\vrule width 1pt}c!{\vrule width 1pt}c!{\vrule width 1pt}c!{\vrule width 1pt}}
\noalign{\hrule height 1pt}
 & $\bra{4,0}$ & $\bra{3,1}$ & $\bra{2,2}$ & $\bra{1,3}$ & $\bra{0,4}$ \\
\noalign{\hrule height 1pt}
$\ket{4,0}$ & $\op{2,0}{2,0}$ & $\op{2,0}{1,1}$ & $\op{2,0}{0,2}$ & $0$ & $0$ \\
\noalign{\hrule height 1pt}
$\ket{3,1}$ & $\op{1,1}{2,0}$ & $\op{1,1}{1,1} + 2\op{2,0}{2,0}$ & $\op{1,1}{0,2} + 2\op{2,0}{1,1}$ & $2\op{2,0}{0,2}$ & $0$ \\
\noalign{\hrule height 1pt}
$\ket{2,2}$ & $\op{0,2}{2,0}$ & $\op{0,2}{1,1} + 2\op{1,1}{2,0}$ & $\op{0,2}{0,2} + 2\op{1,1}{1,1} + \op{2,0}{2,0}$ & $2\op{1,1}{0,2} + 2\op{2,0}{1,1}$ & $\op{2,0}{0,2}$ \\
\noalign{\hrule height 1pt}
$\ket{1,3}$ & $0$ & $2\op{0,2}{2,0}$ & $2\op{0,2}{1,1} + 2\op{1,1}{2,0}$ & $2\op{0,2}{0,2} + \op{1,1}{1,1}$ & $\op{1,1}{0,2}$ \\
\noalign{\hrule height 1pt}
$\ket{0,4}$ & $0$ & $0$ & $\op{0,2}{2,0}$ & $\op{0,2}{1,1}$ & $\op{0,2}{0,2}$ \\
\noalign{\hrule height 1pt}
\end{tabular}
\caption{Transformation of $R^4$  symmetric states to $R^2$ under double partial trace}
\label{tab:R4_to_R2_partial_trace}
\end{table*}

\end{document}